\documentclass[aps,twocolumn,amsmath,amssymb,prl,superscriptaddress,preprintnumbers,nofootinbib,10pt]{revtex4-1}
\usepackage{graphicx}

\newcommand\eqr[1]{(\ref{#1})}
\newcommand\figr[1]{Fig.\,\ref{#1}}
\newcommand\ds{\displaystyle}
\newcommand\ts{\textstyle}

\newcommand\Tr{{\rm Tr}}

\newcommand\del{\partial}

\newcommand\cC{{\cal C}}

\newcommand\cF{{\cal F}}

\newcommand\cL{{\cal L}}

\newcommand\cN{{\cal N}}
\newcommand\cO{{\cal O}}

\newcommand\cT{{\cal T}}

\newcommand\bC{{\mathbb C}}
\newcommand{\be}{\begin{equation}}
\newcommand{\ee}{\end{equation}}
\newcommand{\bea}{\begin{eqnarray}}
\newcommand{\eea}{\end{eqnarray}}

\begin{document}
\preprint{MCTP-11-45}
\preprint{QMUL-PH-11-22}
\title{$AdS$/CFT Dual Pairs from M5-Branes on Riemann Surfaces}
\author{Ibrahima Bah}
\affiliation{Michigan Center for Theoretical Physics, University of Michigan, Ann Arbor, MI 48109, USA}
\author{Christopher Beem}
\author{Nikolay Bobev}
\affiliation{Simons Center for Geometry and Physics, Stony Brook University, Stony Brook, NY 11794-3636, USA}
\author{Brian Wecht}
\affiliation{Center for the Fundamental Laws of Nature, Harvard University, Cambridge, MA 02138, USA}
\affiliation{Centre for Research in String Theory, Queen Mary, University of London, London E1 4NS, UK}
\begin{abstract}
\noindent In this letter we describe an infinite family of new $\cN=1$ $AdS_5$/CFT${}_4$ dual 
pairs which arise from M5-branes wrapping Riemann surfaces in Calabi-Yau threefolds. We use 
the relevant brane constructions to compute the central charges of the infrared fixed points from
the M5-brane anomaly polynomial. We then present $AdS_5\times M_6$ solutions of eleven-dimensional 
supergravity which are dual to these CFTs at large $N$. Finally, we provide a purely four-dimensional 
field theory construction which flows to a special class of these fixed points. These theories are 
further elaborated upon in a companion paper \cite{B3W2}.
\end{abstract}
\maketitle

Despite many years of study, the M5-brane remains a deeply mysterious object. When wrapped on a complex curve $\cC_g$ of genus $g>1$ inside a Calabi-Yau manifold, M5-branes support a low-energy four-dimensional worldvolume theory that is strongly coupled and generically has no known Lagrangian description. In certain cases, explicit solutions of eleven-dimensional supergravity are known which provide a holographic description of such theories at large $N$, and even at finite $N$ some aspects of these theories can be understood. Still, the set of such solutions is small, and the number of known field theory duals even smaller.

The first example of such eleven-dimensional solutions was discovered by Maldacena and N\'u\~nez (MN) \cite{Maldacena:2000mw}. These solutions preserve either four or eight supercharges, and the near-horizon geometry has an $AdS_5$ factor. The dual field theories are therefore four-dimensional $\cN=1$ or $\cN=2$ SCFTs. Although for many years these theories remained mysterious, the recent discovery \cite{Gaiotto:2009we} of a new class of $\cN=2$ SCFTs provided the missing link. These new theories, which are called $T_N$, have $SU(N)^3$ global symmetry and are strongly coupled. In \cite{Gaiotto:2009gz}, the authors showed that the $\cN=2$ MN solutions are dual to $T_N$ theories coupled together by pairing all non-Abelian global symmetries and gauging the diagonal subgroups. Furthermore, in \cite{Benini:2009mz}, it was shown that the $\cN=1$ MN solutions are dual to Gaiotto's theories after perturbing by a mass term for all the adjoint chiral superfields in $\cN=2$ vector multiplets and flowing to the infrared.

It is useful to organize these constructions in terms of the ambient Calabi-Yau geometry in which the curve $\cC_g$ is embedded (in this note, we consider only the case where $\cC_g$ is closed). In theories preserving eight supercharges, this ambient space is a Calabi-Yau twofold, and locally the  geometry is the cotangent bundle $\cT^\star\cC_g$. This is the arrangement which leads to the $\cN=2$ MN solutions. Alternatively, when four supercharges are preserved, the ambient space is a Calabi-Yau threefold. The $\cN=1$ MN solutions then result from the situation in which the local geometry is of the form $K^{1/2}\oplus K^{1/2}\rightarrow\cC_g$ with $K$ the canonical bundle of the curve.

In this work, we establish an infinite family of new eleven-dimensional solutions and their field theory duals which arise on M5-branes wrapping complex curves $\cC_g$ in Calabi-Yau threefolds. In contrast to the situations described above, we allow the local geometry of the threefold to be any decomposable $\mathbb{C}^2$-bundle over $\cC_g$. The bundle then splits as a sum of line bundles $\cL_1\oplus\cL_2\rightarrow\cC_{g}$. The Calabi-Yau condition requires $\cL_1\otimes\cL_2=K$. Taking $p=\deg\cL_1$ and $q=\deg\cL_2$, this implies that $p+q=2g-2$. Any such choice of $p$ and $q$ leads to an allowed twist of the M5-brane theory on $\cC_{g}$ preserving at least four supercharges. For $q=0$ or $p=0$, there is an enhancement to eight supercharges and the construction reduces to that of \cite{Gaiotto:2009we}. For $p=q$ we recover the $\cN=1$ theories studied in \cite{Benini:2009mz}. 

\medskip\noindent {\bf New fixed points from six dimensions.} While very little can be computed directly from the six-dimensional theory of M5-branes, it is possible to extract the superconformal R-symmetry and central charges of the putative four-dimensional IR fixed points by an anomaly computation similar to the one performed in \cite{Benini:2009mz}. This will provide a stringent check on our interpretation of the gravity and field theory constructions which follow. 

The anomaly eight-form for $N$ M5-branes is \cite{Harvey:2005it}
\be 
I[N]=(N-1) I[1]+(N^3-N)\frac{p_2(\cN)}{24}~,
\label{NM5anomaly}
\ee
where $I[1]$ is the anomaly for a single M5-brane,
\be 
I[1]=\tfrac{1}{48}[p_2(\cN)-p_2(\cT)+\tfrac{1}{4}(p_1(\cT)-p_1(\cN))^2]~.
\label{oneM5anomaly}
\ee
The characteristic classes $p_{1,2}$ appearing in these expressions are the first and second Pontryagin classes of the normal bundle $\cN$ and tangent bundle $\cT$ to the six-dimensional M5-brane worldvolume. The choice of normal geometry implies that the Chern roots $\ell_1, \ell_2$ of the normal bundle are related to the Chern class $t$ of the tangent bundle to the Riemann surface by
\be \ell_1 = - \frac{1+z}{2} t~,\quad \ell_2 = - \frac{1-z}{2} t~.\label{normalbundles}\ee
The parameter $z$ is related to the degrees of the line bundles $\cL_{1,2}$ by $z=(p-q)/(2g-2)$.

The central charges of the resulting four-dimensional theory can be obtained by turning on a background flux for the superconformal R-symmetry and then integrating the anomaly eight-form over $\cC_g$ to produce an anomaly six-form. However, the decomposable nature of the $\bC^2$ bundle over $\cC_g$ implies that the field theory has a $U(1)^2$ symmetry coming from independent rotations of the line bundles $\cL_{1,2}$, so the superconformal R-symmetry is not {\it a priori} identifiable.

It is useful to define subgroups of this $U(1)^2$ symmetry as follows: $U(1)_{R_0}$, which rotates the two normal line bundles with equal phases, and $U(1)_{\cF}$, which rotates them with opposite phases and under which the preserved supercharges of the twisted M5-brane theory are invariant. We can then parameterize our ignorance of the infrared superconformal R-symmetry in terms of a trial R-symmetry $R_{\epsilon}=R_0+\epsilon\cF$. Coupling $R_\epsilon$ to a $U(1)$ bundle $F$ induces a shift in the Chern roots,
\be 
\ell_1 \rightarrow \ell_1 + (1+\epsilon) c_1(F)~,~~~\ell_2 \rightarrow \ell_2 + (1-\epsilon) c_1(F)~.
\label{shifts}
\ee
The anomaly six-form then takes the form
\be 
I_6 = \tfrac{1}{6}{\rm Tr} R_{\epsilon}^3 c_1 (F)^3 -\tfrac{1}{24} {\rm Tr} R_{\epsilon} \, c_1 (F) p_1 (\cT_{4})~,
\label{sixform}
\ee
in terms of the linear and cubic 't Hooft anomalies of the trial R-symmetry. The central charges $a$ and $c$ are linear combinations of these anomalies when $R_{\epsilon}$ is the superconformal R-symmetry, with the value of $\epsilon$ fixed by the requirement that $a$ be maximized \cite{Intriligator:2003jj}. 

Explicit expressions for the maximizing value of $\epsilon$ and the central charges are unwieldy and appear in \cite{B3W2}. However, the answers simplify at large $N$, where we find that for the superconformal R-symmetry,
\be 
\epsilon = \epsilon_* \equiv \frac{1 - \sqrt{1 + 3 z^2}}{3z}~,
\label{epsilondef}
\ee
and the central charges are given by
\be 
a(z) = c(z) = \ds\frac{(g-1)N^3}{48z^2} ( (1+3 z^2)^{3/2} + 9z^2-1 )~.
\label{centralcharge}
\ee

\begin{figure}
\includegraphics[width=7.6cm]{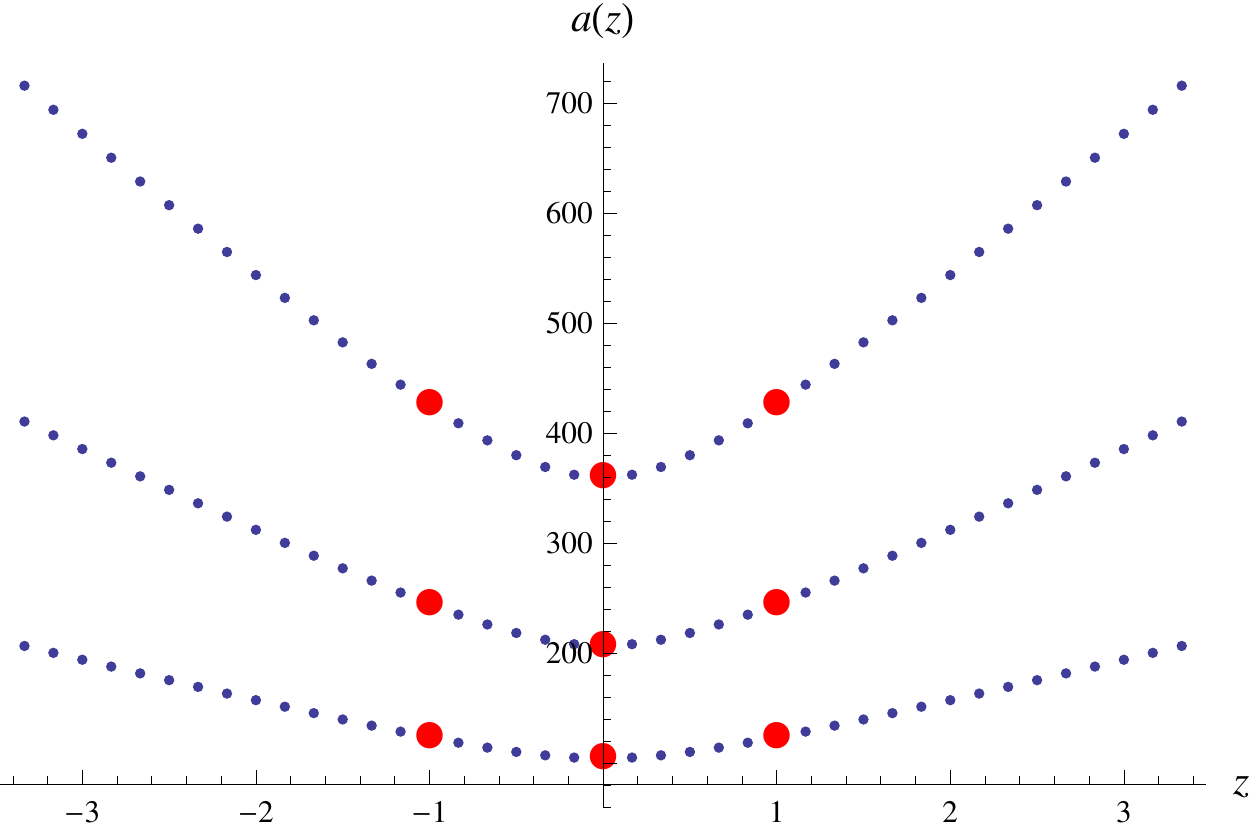}
\caption{\label{ccplot} The central charge $a$ as a function of the twist parameter $z$ for genus $g=7$ and $N=4,5,6$ (bottom to top). The MN theories are marked with large points at $z=0$ and $|z|=1$.}
\end{figure}

The organization of the fixed points for a given $\cC_g$ reflects the structure of the normal geometry to the M5-branes (see \figr{ccplot}). In particular, there are $g-2$ distinct fixed points with central charges that fall between those of the $\cN=1$ and $\cN=2$ MN twists, corresponding to line bundles with $p$ and $q$ both positive. There are then infinitely many fixed points with central charge greater than that of the $\cN=2$ MN twists, with the central charges growing linearly for large $z$.

\medskip\noindent {\bf Dual supergravity backgrounds.} At large $N$, the existence of the fixed points described above can be established holographically. The dual supergravity solutions can be found by imposing constraints on the eleven-dimensional spinors to match the conserved supersymmetries on the M5-branes. The supergravity solutions thus correspond to ``wrapped-brane spacetimes'' in the language of \cite{Gauntlett:2006ux}. The resulting backgrounds admit a truncated description in terms of a $U(1)^2$ seven-dimensional gauged supergravity, as was the case for the special twists studied in \cite{Maldacena:2000mw}. In fact, BPS flows from an asymptotically locally $AdS_7$ geometry to the $AdS_5$ fixed points can be found numerically,  demonstrating that RG flows from the twisted six-dimensional field theories actually hit the four-dimensional fixed points in question \cite{B3W2}.

We find a two-parameter family of eleven-dimensional $AdS_5$ solutions indexed by $g$ and $z$ that can be obtained from the seven-dimensional solutions by the uplift formul{\ae} of \cite{Cvetic:1999xp}. (A subset of our solutions has previously appeared in \cite{Cucu:2003yk}.) The metric takes the form
\be
\begin{split}
&ds^2_{11}= {\Delta}^{1/3} ds_7^2 +\ts\frac{1}{4}\Delta^{-2/3}ds^2_{4}~,\\
&ds_7^2= e^{2f_0}\ds\frac{-dt^2+d\vec{z}^{\,2}+dr^2}{r^2} + e^{2g_0} \ds\frac{dx^2+dy^2}{y^2}~,\\
&ds^2_4=X_0^{-1} d\mu_0^2 +\ds\sum_{i=1}^{2}X_{i}^{-1} (d\mu_i^2+\mu_i^2(d\phi_i+A^{(i)})^2)~,
\label{11dsolution}
\end{split}
\ee
where we have defined
\be
\nonumber\Delta = \ds\sum_{\alpha=0}^{2} X_{\alpha} \mu_{\alpha}^2~, ~~ \ds\sum_{a=0}^{2}\mu_a^2 = 1~, ~~ A^{(i)} = a_i \ds\frac{dx}{y}\pm A_{\it flat}~,
\ee
with $a_1\equiv (1+z)/2$, $a_2\equiv (1-z)/2$, and $A_{\it flat}$ the connection for a flat $S^1$ bundle over $\cC_g[x,y]$. The constants take values depending on $z$,
\bea
&&X_1^5= \ds\frac{1+7z+7z^2+33z^3-(1+4\,z+19\,z^2)\sqrt{1+3\,z^2}}{4\,z(1-\,z)^2}~, \notag\\
&& X_1X_2^{-1}= \ds\frac{1+\,z}{2\,z+ \sqrt{1+3\,z^2}}~, ~~ X_0 = \left(X_1X_2\right)^{-2}~,\\
&&e^{f_0} = X_0^{-1}~,~~e^{2g_0} =  \, \ds\frac{X_1X_2}{8} \left((1-z)X_1+ (1+z) X_2\right)~.\notag
\eea
The four-form flux is given by
\be
\begin{split}
&*_{11}F_{(4)}= 4\ds\sum_{\alpha=0}^{2} (X_{\alpha}^2 \mu_{\alpha}^2 -\Delta X_{\alpha})\epsilon_{(7)} + 2 \Delta X_{0} \epsilon_{(7)}\\
&\qquad+\ts\frac{1}{16} \ds\sum_{i=1}^{2} X_{i}^{-2} d(\mu_i^2) \wedge (d\phi_i+A^{(i)}) \wedge *_{7} F^{(i)}~,
\label{flux}
\end{split}
\ee
where $F^{(i)}=dA^{(i)}$, $\epsilon_{(7)}$ is the volume form for $ds^2_7$, and $*_7$ and $*_{11}$ denote the Hodge star operator with respect to $ds^2_{7}$ and $ds^2_{11}$, respectively.

The internal manifold in the metric \eqr{11dsolution} is an $S^4$ fibration over $\cC_{g}$, which is realized as an appropriate quotient of $\mathbb{H}^2$. Compatibility of the $S^4$ fibration with this quotient then requires that $z$ take discrete values as in the previous section. This structure exactly parallels the geometry of the M5-brane construction. One can identify the Killing vector generating $U(1)_{\cF}$ with $\partial_{\phi_1}-\partial_{\phi_2}$ and the one corresponding to $U(1)_{R_0}$ with $\partial_{\phi_1}+\partial_{\phi_2}$. It is also apparent that there are exactly marginal deformations of these backgrounds given by variations of the complex structure of the curve $\cC_g$ (realized by modifying the quotient action on $\mathbb{H}^2$) along with variations of the flat connection $A_{\it flat}$. The conformal manifold of the dual field theories is then of complex dimension $(3g-3)+g$.

It is straightforward to compute the central charges for these backgrounds using holography, and the result is in precise agreement with \eqr{centralcharge}. Furthermore, there are BPS operators $\mathcal{O}_{M2}$ in these theories which are dual to M2-branes wrapping the curve $\cC_{g}$. The dimension of such an operator is also computable from the gravity solution \cite{Gaiotto:2009gz}, and is given by
\be 
\Delta[\mathcal{O}_{M2}]=N(g-1)\left(1+\ts\frac{1}{2}\sqrt{1+3z^2}\right).
\label{gravm2}
\ee

A canonical system of coordinates for supersymmetric $AdS_5$ solutions of eleven-dimensional supergravity was introduced in \cite{Gauntlett:2004zh}. A prominent role in this construction is played by a particular Killing vector -- constructed from a constant-norm Killing spinor -- which generates the superconformal R-symmetry of the dual field theory. The solutions in \eqr{11dsolution} can be adapted to these coordinates, in which case the relation of the superconformal Killing vector to the two Killing vectors in \eqr{11dsolution} is given by
\be \del_\psi = \ts\frac{4}{3}X_1^2X_2^2\left(X_1\del_{\phi_1}+X_2\del_{\phi_2}\right)~.\ee
This geometric identification of the superconformal R-symmetry matches the result of $a$-maximization in the previous section.

\medskip\noindent {\bf Field theory construction.} We now turn to the field theories dual to the above constructions. The basic building blocks will be the $T_N$ theories discovered by Gaiotto in \cite{Gaiotto:2009we}. These are isolated $\cN=2$ SCFTs with $SU(N)^3$ global symmetry coming from $N$ M5-branes wrapping a thrice-punctured sphere $\cC_{0,3}$. Although there is no known weakly coupled description of these objects for $N>2$, we do know some of the gauge-invariant operators and their dimensions. In particular, each $T_N$ comes with three dimension-two operators $\mu_a$, $a = 1,2,3$, each transforming in the adjoint of one of the $SU(N)$ groups. There are also operators $Q, \widetilde Q$ of dimension $N-1$, transforming in the $({\bf N},{\bf N},{\bf N})$ and $(\overline {\bf N},\overline{\bf N},\overline{\bf N})$ representations, respectively.

Diagrammatically, we represent a $T_N$ by a triangle (trinion) with three prongs coming off the vertices, corresponding to the three $SU(N)$ global symmetries. One can now construct a large family of new $\cN=2$ field theories by gauging the diagonal subgroup of two of these $SU(N)$'s, generically from different $T_N$ blocks. To be consistent with $\cN=2$ supersymmetry, we need to include the superpotential term $W= \mbox{Tr}(\mu\phi)$, where $\phi$ is the adjoint chiral superfield in the vector multiplet. These theories have $p$ or $q$ equal to zero and are dual to the $\cN=2$ MN solutions.
Similarly, one could connect the $T_N$ blocks with all $\cN=1$ vector multiplets, as discussed in \cite{Bah:2011je}. The resulting theories then have $p=q$ and are dual to the $\cN=1$ MN solutions.

Coupling $T_N$ blocks with vector multiplets has a geometric interpretation in terms of gluing together local Calabi-Yau threefold geometries of the form ${\mathbb C}\times\cT^*\cC_{0,3}$. In the $\cN=2$ case, the line bundles are connected so that the resulting geometry is still the sum of a trivial bundle and a cotangent bundle. Alternatively, we interpret coupling by an $\cN=1$ vector as gluing the local geometries so that the cotangent bundle over one sphere is connected to the trivial bundle over the other and vice versa. For any one of our local geometries with $p$ and $q$ both non-negative, we can introduce a decomposition into a collection of $2g-2$ punctured spheres, over each of which is fibered one curved and one trivial line bundle. Furthermore, each punctured sphere is endowed with a number $k_i=1,2$ indicating that the curved part of the normal geometry lies in the line bundle $\cL_{k_i}$.

\begin{figure}
\includegraphics[width=6.0cm]{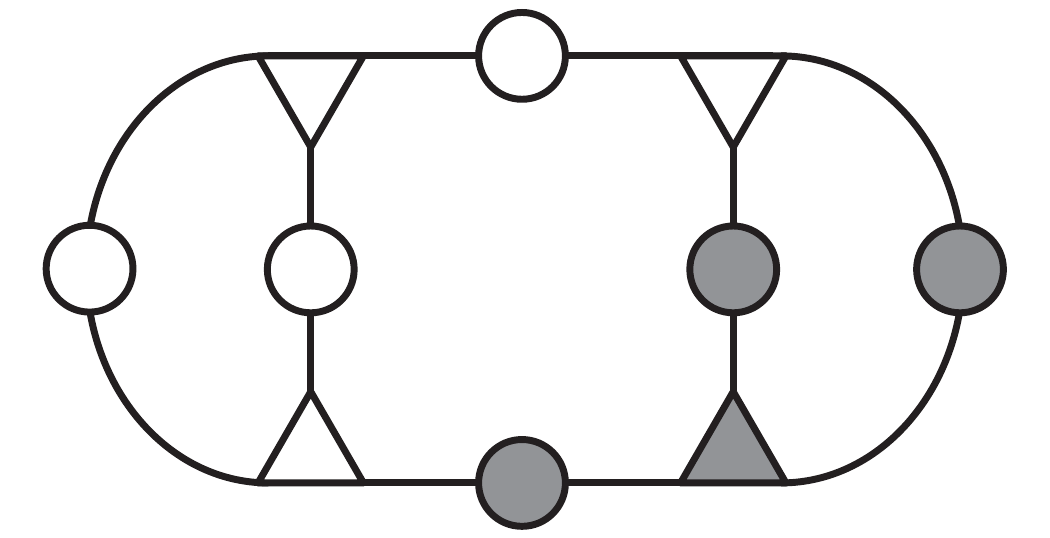}
\caption{\label{genus3} An example of an $\cN=1$ quiver construction at genus $g=3$ with $(p,q)=(1,3)$. Shaded trinions have $\sigma=+1$ while unshaded have $\sigma=-1$. Shaded versus unshaded nodes represent $\cN=1$ and $\cN=2$ gluings, respectively.}
\end{figure}

To each decomposition of this kind, we associate a generalized $\cN=1$ quiver theory where the $i$'th $T_N$ block in the quiver comes with a sign $\sigma_i=(-1)^{k_i+1}$, and the connecting gauge fields come in $\cN=2$ or $\cN=1$ vector multiplets according to whether they connect two $T_N$ blocks with $\sigma_i\sigma_j=+1$ or $-1$, respectively. There will be $p$ blocks with $\sigma_i =+1$ and $q$ with $\sigma_i=-1$. Any such arrangement is naturally encoded in a generalized quiver diagram with shading (see \figr{genus3}). In such a construction the $\cN=1$ gauge interactions are asymptotically free, and in the UV the theory is a collection of $\cN=2$ generalized quiver theories along with some decoupled $\cN=1$ vector multiplets. 

Our main tool to analyze these field theories will be global symmetries. When viewed as an $\cN=1$ object, each $T_N$ has two global $U(1)$ symmetries, $R_{\cN=1,i}$ and $J_i$, which are in the Cartan subalgebra of the $SU(2)\times U(1)$ $\cN=2$ R-symmetry. The charges of these symmetries are as described in \cite{Benini:2009mz}. Each $\cN=2$ vector multiplet additionally comes with an adjoint chiral superfield charged under a global symmetry $F_{i_a j_b}$. The indices $i_a$ and $j_b$, with $a,b = 1,2,3$, refer to the possible $SU(N)$ flavor groups of the two $T_N$'s attached to the vector multiplet.

When all the $T_N$ theories are connected with either $\cN=1$ or $\cN=2$ vector multiplets, there is one anomaly-free R-symmetry $R_0$ and one anomaly-free non-R symmetry $\cF$. Given any choice of $\{ \sigma_i \}$, $R_0$ and $\cF$ are
\be
\nonumber R_0 = R+\tfrac{1}{6}\sum_i J_i~,~~ \cF = \tfrac 12\sum_i \sigma_i J_i +\tfrac 12 \sum (\sigma_i+\sigma_j)F_{i_aj_b}~.
\ee
The last sum is over all locations of $\cN=2$ vector multiplets, {\it i.e.}, between $T_N$ blocks $i$ and $j$ with $\sigma_i \sigma_j = 1$. $R$ is the superconformal R-symmetry of the UV fixed point, and acts as $R_{\cN=1,i}$ on the $i$'th $T_N$. The symmetries $R_0$ and $\cF$ then admit a natural interpretation as the symmetric and anti-symmetric combinations of the $U(1)$ actions on the line bundles $\cL_1$ and $\cL_2$. 

Since $R_0$ and $\cF$ can potentially mix, there is a one-parameter family of R-symmetries given by $R_\epsilon=R_0+\epsilon \cF$. The superconformal R-symmetry can then be fixed by maximizing $a_\epsilon = \frac{3}{32} \left (3 {\rm Tr} R_{\epsilon}^3 - {\rm Tr} R_{\epsilon} \right )$. The answer is in precise agreement with the calculation using the previous methods.

At large $N$ the central charge $a_\epsilon$ is maximized at $\epsilon = \epsilon_*$ \eqr{epsilondef} (the more unwieldy value of $\epsilon$ also matches the anomaly analysis for finite $N$, see \cite{B3W2}). The superconformal R-charges of some of the chiral primary operators are then
\be\begin{split}
&R[\mu_{i}] = 1- \sigma_i \epsilon_*~,\\
&R[\phi_{ij}] = 1 +\ts\frac{1}{2} (\sigma_i + \sigma_j) \epsilon_*~,\\
&R[Q_i]=R[\widetilde{Q}_i]=\ts\frac12 (N-1)(1-\sigma_i \epsilon_*)~.
\end{split}\ee
The dimension-three chiral operators at such a fixed point are easily enumerated. For each $\cN=2$ node of the quiver, there are operators $\Tr(\mu_{i_a}\phi_{i_a j_b})$ and $\Tr(\mu_{j_b}\phi_{i_a j_b})$, while for each $\cN=1$ node there is a single operator $\Tr(\mu_{i_a}\mu_{j_b})$. To compute the dimension of the conformal manifold, we can use the methods of Leigh and Strassler \cite{Leigh:1995ep}.  In a theory with $m_2$ $\cN=2$ vector multiplets and $m_1$ $\cN=1$ vector multiplets, there are $2m_2 + m_1$ marginal operators. Adding this to $3g-3$ gauge couplings, we obtain $(3g-3)+2m_2+m_1$ marginal parameters. However, $(2g-3)+m_2$ of these are removed by anomaly constraints and field redefinitions, leaving $(3g-3) + g$ exactly marginal parameters. This matches the gravity calculation of the complex dimension of the conformal manifold. Finally, we note that an M2-brane wrapping the Riemann surface is dual to $\cO_{M2}=\prod_i Q_i$, with dimension $\ts\frac32(g-1)(N-1)(1-\epsilon_* z)$, in agreement with \eqr{gravm2}.

\medskip\noindent {\bf Acknowledgements} We would like to thank F.~Benini, M.~Douglas, J.~Gauntlett, J.~Maldacena, D.~Morrison, L.~Rastelli, B.~van Rees, D.~Waldram and E.~Witten for useful discussions. We also thank the Simons Workshop in Mathematics and Physics 2011 for a stimulating working environment during the initial stages of this project. The work of CB and NB is supported in part by DOE grant DE-FG02-92ER-40697. IB supported by DOE grant DE-FG02-95ER-40899. BW is supported by the Fundamental Laws Initiative of the Center for the Fundamental Laws of Nature, Harvard University, and the STFC Standard Grant ST/J000469/1 ``String Theory, Gauge Theory and Duality."

\bibliography{class_s_prl}
\end{document}